\documentstyle[12pt]{article}
\input epsf.sty 
\newcommand{\NP}[1]{ Nucl.\ Phys.\ {\bf #1}}

\newcommand{\PL}[1]{ Phys.\ Lett.\ {\bf #1}}

\newcommand{\PR}[1]{Phys.\ Rev.\ {\bf #1}}
\newcommand{\PRL}[1]{ Phys.\ Rev.\ Lett.\ {\bf #1}}

\newcommand{\tr}{\mbox{tr}}

\newcommand{\nn}{\nonumber}

\newcommand{\Od}{{\cal O}}
\newcommand{\Opd}{{\cal O}(p^2)}
\newcommand{\Opc}{{\cal O}(p^4)}

\newcommand{\be}{\begin{equation}}
\newcommand{\ee}{\end{equation}}
\newcommand{\bea}{\begin{eqnarray}}
\newcommand{\eea}{\end{eqnarray}}
\newcommand{\qqo}{\langle0\vert \bar q q\vert 0 \rangle}
\newcommand{\mh}{\hat{m}}

\begin{document}
\thispagestyle{empty}
\begin{titlepage}
\title {\vspace{-2.0cm}
\hfill {\normalsize SLAC-PUB-7881} \\ \vspace{-0.3cm}
\hfill {\normalsize hep-ph/9807337} \\ \vspace{0.6cm}}

\title{ On the Size of the Chiral Condensate\\ 
Generalized Chiral Perturbation Theory\\
and\\
the DIRAC experiment\footnote{Research partially supported
by the Department of Energy under contract DE-AC03-76SF00515}} 
\author{
A. Dobado\footnote{E-mail: dobado@eucmax.sim.ucm.es}\\
{\small\em Departamento de F\'{\i}sica Te\'orica.}\\
{\small\em Universidad Complutense. 28040 Madrid. Spain.}\\
\\
J. R. Pel\'aez\footnote{E-mail:pelaez@slac.stanford.edu}
\footnote{On 
leave of absence from the Departamento de F\'{\i}sica Te\'orica.
Universidad Complutense. 28040 Madrid. Spain.}\\
{\small\em Stanford Linear Accelerator Center}\\
{\small\em Stanford University, 
Stanford, California 94309. U.S.A.}}

\maketitle   
\thispagestyle{empty}
\begin{abstract}
 In the near future, the DIRAC collaboration will measure
$\pi\pi$ scattering lengths with great precision. Those measurements
are likely to shed some light on the problem of the size of the chiral
$\qqo$ condensate.  Although it is usually assumed to be as a large as
$\sim(-225\,\mbox{MeV} )^3$, in the last years a more general approach,
has been developed to
accommodate either a large or a small alternative $\sim(-100
\,\mbox{MeV})^3$. Such a low value would also modify the 
standard temperature estimate at
which the chiral phase transition occurs. In this work we briefly
review the basic theoretical ideas related to this issue as well as
the experiment that could help to establish any of the two scenarios. 
\end{abstract}

\begin{center}
{\it Invited contribution to the ``Gribov Memorial Volume''}

{\it Heavy Ion Physics. Acta Physica Hungarica. New Series.}
\end{center}
\end{titlepage}

\setcounter{page}{1} 

\section{Introduction} 
In spite of the common belief that QCD is the appropriate theory
describing strong interactions, it is very few what this theory can
tell us about low energy hadron physics. In this regime the main
guiding fact is the chiral symmetry $SU(3)_L \times SU(3)_R$ of the
QCD Lagrangian in the massless quark limit ($m_u=m_d=m_s=0$). For
still not quite well understood dynamical reasons, this global
symmetry is spontaneously broken to the diagonal group
$SU(3)_{L+R}$. Thus the eight lowest mass $J^P=0^-$ mesons $\pi,K$ and
$\eta$ emerge as the corresponding Goldstone bosons. Indeed, their
relatively low masses can be understood from the fact that quarks do
actually have some small masses which, even though they modify
the previous picture, still can be treated as small
perturbations. Using chiral symmetry and these small quark masses it
was possible in the early sixties to derive many important relations
between hadron masses \cite{GOR}. For example, in the exact isospin
limit ($m_u=m_d$) it was proposed the celebrated Gell-Mann-Okubo mass
formula
\be
4M_K^2-M_\pi^2-3M_\eta^2\simeq0\qquad
\ee
Associated with the spontaneous chiral symmetry breaking is the
so-called quark condensate $\qqo$. A non-vanishing value of this
magnitude breaks the $SU(3)_L \times SU(3)_R$ chiral symmetry, but not
the diagonal $SU(3)_{L+R}$ group. Thus it is a good candidate to play
the role of the order parameter associated with the chiral symmetry
broken phase. At higher temperatures, where the chiral symmetry is
expected to be restored, the thermal condensate, $\qqo_T$, should
vanish above some critical temperature $T_c$. This chiral phase
transition, together with the deconfinement transition, has raised a
considerable interest, both theoretically and experimentally, in the
heavy ion physics community. In this paper we will review recent works
that have proposed an alternative to the standard scenario we have
just described. Indeed, we will discuss about the precise value of the
quark condensate, its relation with other low energy hadron parameters
and the possibly of measuring it in future experiments. The plan of
the paper is as follows. In Sec.2 we make some introductory remarks on
the role of the condensate and spontaneous chiral symmetry
breaking. In Sec.3 we review the effective Lagrangian formalism. In
Sec.4 we introduce the explicit symmetry breaking mass terms and
compare Chiral Perturbation Theory with the so called Generalized
Chiral Perturbation Theory. In Sec.5 we review the present
experimental evidence and the predictions in the different
scenarios. In Sec.6 we deal with the chiral condensate evolution with
the temperature, which could change if the standard picture has to be
modified. In Sec.7 we describe briefly the DIRAC experiment where
pionium ground states will be produced in order to get very precise
measurements of the pion scattering lengths that could provide
decisive information about the quark condensate. Finally in Sec.8 we
summarize.

\section{The chiral condensate} 

As it was commented above, in the standard scenario it is assumed that
the quark condensate is the order parameter of the
chiral phase transition \cite{GOR}. In the chiral limit
($m_u=m_d=m_s=0$) it is customary to define the constant

\be
B_0\equiv-\frac{\langle 0\vert \bar{u} u\vert 0 \rangle}{F^2}\label{B_0}
=-\frac{\langle 0\vert \bar{d} d\vert 0 \rangle}{F^2}
=-\frac{\langle 0\vert \bar{s} s\vert 0 \rangle}{F^2}
\label{defB_0}
\ee
where $F\simeq 90 \,\mbox{MeV}$ is the pion decay constant. Thus the standard
wisdom can be simply summarized by the {\em assumption} that 
$2\hat m B_0$ (with $2 \hat m = m_u+m_d$) is roughly the pion
mass squared $M_{\pi}^2$, so that $B_0\simeq 1.3\,
\mbox{GeV}$. The resulting condensate is $\qqo\simeq(-225\,
\mbox{MeV})^3$. In this scenario the critical temperature $T_c$ is
expected to be around $200 \,\mbox{MeV}$ for two flavors
\cite{GeLe,QSR}. This large condensate assumption leads naturally to
the Gell-Mann-Okubo formula. The effect of the quark masses can be
included systematically in this scheme giving rise to the
standard Chiral Perturbation Theory (ChPT) \cite{We,GaLe}. 

However, in
spite of the above argument in favor of the standard
scenario, the Gell-Mann-Okubo formula can hold quite independently of
the relation between $2\hat m B_0$ and $M_{\pi}^2$. In fact, an
alternative completely consistent formalism where $2\hat m B_0$ is
considerably lower than $M_{\pi}^2$ has been recently proposed
\cite{GChPT1}. In the new scheme the quark condensate could
be much smaller than in the standard case, typically
around $100\, \mbox{MeV}$. The corresponding perturbative treatment of
the quark mass effects is called Generalized Chiral Perturbation
Theory (GChPT). 

There is a very illustrative analogy of the above two
alternatives with spin systems \cite{Stern}. Whenever spins are
strongly correlated with their nearest neighbors we have an ordered
phase at low temperatures and a disordered one above some critical
temperature $T_c$. In a system with just one kind of spins we have two
possible cases. On the one hand, if the exchange coupling constant $J$
is positive, the interaction will favor parallel spins. Even in the
absence of any external magnetic field, in the ordered phase the spins
will be nearly parallel, thus yielding a macroscopic magnetization,
which is therefore a good order parameter to distinguish between the
ordered and the disordered phase. Such a system is nothing but a
ferromagnet and $T_c$ is called the Curie temperature. That would be
analogous to the standard ChPT case, where the magnetization would
play the role of the quark condensate and the quark masses would play
the role of some external magnetic field.  (Note however that the
analogy cannot be carried too far since, among other things, the
symmetry groups are different). On the other hand, when $J$ is
negative, and even if $T<T_c$ the spins are nearly antiparallel and no
macroscopic magnetization is produced. Thus the magnetization is not a
good order parameter to distinguish between the two phases, the system
is antiferromagnetic and $T_c$ is called the N\'eel temperature. That
would be similar to the extreme case of GChPT where $B_0=0$. Note that
general principles do not exclude the possibility that the quark
condensate vanishes in the chiral limit. Another interesting
possibility appears when different kinds of spins are
considered. Then, even for $J<0$ it is possible to generate a
macroscopic magnetization in the antiparallel ordered state for
$T<T_c$ and thus the magnetization is still a good order
parameter. This kind of systems are called ferrimagnets (natural
magnets are of this type). This case would be the most general to
establish the analogy with QCD.

\section{Effective Lagrangians}
 We have just seen that pions, kaons and etas can be identified with
 Goldstone bosons and thus we will parameterize them in an $SU(3)$
 matrix as follows: 
\be 
U=\exp(i\Phi/F)\quad ; \quad
 \Phi=\sqrt{2}\left( 
\begin{array}{ccc}
 \frac{1}{\sqrt{2}}\pi^0+\frac{1}{\sqrt{6}}\eta&\pi^+&K^+\\
 \pi^-&-\frac{1}{\sqrt{2}}\pi^0+\frac{1}{\sqrt{6}}\eta&K^0\\
 K^-&\bar{K}^0&-\frac{2}{\sqrt{6}}\eta\\ 
\end{array} \right) 
\ee 
Let
 us consider first the chiral limit. In this case the GB are massless
 and therefore they are the most relevant degrees of freedom at
 sufficiently low energies or momenta.  The philosophy of Chiral
 Perturbation Theory \cite{GaLe} (see also ref.\cite{libro}) is
 to perform a low momentum expansion, or what it is the same, an
 expansion in the number of derivatives in the
 Lagrangian. Generically, low momenta means much smaller than the
 typical hadronic scale of around $\Od(1\,\mbox{GeV})$.  Thus, the lowest
 order (denoted $\Opd$) Lagrangian is \be {\cal
 L}^{(2)}_{m_q=0}=\frac{F^2}{4}\tr\left( D_\mu UD^\mu U^\dagger\right)
 \ee Note that we have introduced a covariant derivative $D_\mu U=
 \partial_\mu U-ir_\mu U+iUl_\mu$, where $r_\mu$ and $l_\mu$ are right
 and left gauge fields. In order to include the weak and
 electromagnetic interactions of mesons, they can be substituted by
 the corresponding gauge bosons of the Standard Model. If we work out
 the tree level amplitude for, say, $\pi^+\pi^-\rightarrow\pi^0\pi^0$ 
scattering
 we obtain:

\be 
A(s,t,u)=\frac{s}{F^2} \qquad (m_q=0) 
\ee 
From that expression it is possible to obtain all other
$\pi\pi\rightarrow\pi\pi$ scattering processes using isospin and
crossing symmetry. Similar expressions can be found for other
processes involving other mesons, as, for instance, the $I=3/2$
amplitude for elastic $\pi K$ scattering:

\be 
T^{3/2}(s,t,u)=\frac{s}{2F^2}\qquad (m_q=0) 
\ee 
There are two relevant observations common to the above two and other
low order light meson-meson amplitudes: first, that at $s=0$ the interaction
vanishes due to the GB nature of these light mesons. The second
is that there is only one parameter, the scale $F$ in the Lagrangian,
which means that its predictions are the same for any fundamental theory
whose symmetry breaking pattern is the same as in the three flavor QCD
chiral limit.  As a consequence any other specific feature from QCD, apart
from its spontaneous chiral symmetry breaking, shows up at higher
orders. Indeed, at $\Opc$ the most general Lagrangian satisfying
the symmetry constraints is given by:

\bea
 {\cal L}^{(4)}_{m_q=0}&=&L_1\,(\tr(D^\mu U^\dagger D_\mu U))^2+
 L_2\,\tr(D_\mu U^\dagger D_\nu U)\tr(D^\mu U^\dagger D^\nu U) \nn\\
 &+&L_3\,\tr(D^\mu U^\dagger D_\mu U D^\nu U^\dagger D_\nu U)
 -iL_9\,\tr(F^{R}_{\mu\nu}D^\mu U D^\nu U^\dagger+ F^{L}_{\mu\nu}D^\mu
 U^\dagger D^\nu U)\nn\\ &+&L_{10}\,\tr( U^\dagger F^{R}_{\mu\nu} U
 F^{L\,\mu\nu})
 +H_1\tr(F^{R}_{\mu\nu}F^{R\,\mu\nu}+F^{L}_{\mu\nu}F^{L\,\mu\nu})
\eea
 where $F^{R}$ and $F^{L}$ are the usual strength tensors of the
 $r_\mu$ and $l_\mu$ gauge fields, respectively.  With the two above
 Lagrangians it is possible to calculate amplitudes for many different
 processes. Then, comparing with data from just a few processes, we 
can extract the values of the $L_i$ parameters, which
can be used later to obtain predictions for other processes. These
 constants are not fixed just by the scale and do depend on the
 specific dynamics of QCD. However, their actual values cannot be
 computed directly from perturbative QCD, which is not valid at these
 low energies.  Maybe the most relevant fact about these effective
 Lagrangians is that, even though they are not strictly
 renormalizable, it is still possible to calculate loops and obtain
 finite results. For instance, if we calculate diagrams with ${\cal
 L}^{(2)}$ and one loop, it will be possible to absorb all the
 infinities in renormalized $L_i^r$ parameters.  This process can be
 carried on to arbitrarily high orders, and it ensures that, up to a
 given order in momenta, we will always get a finite result.

\section{Quark masses and Explicit Symmetry Breaking}

  Let us now turn on the quark masses. In so doing we are breaking
  {\em explicitly} the Chiral Symmetry, and, consequently, our GB will
  become massive pseudo-GB. Nevertheless, the masses of pions, kaons
  ant etas are still smaller than the typical hadronic scale. We can
  therefore treat them perturbatively and include mass terms in
  the Lagrangian. In addition, we will be assuming exact isospin
  symmetry ($m_u=m_d$).  Since the meson mass matrix term in the
  Lagrangian should vanish when so does the quark mass matrix ${\cal
  M}=diag(\hat m,\hat m, m_s)$, we can write, for instance:
\be 
M^2_{\pi}= 2 B_0 {\hat m} + 4 A {\hat m}^2 + ... \label{masaycondensado} 
\ee 
where $B_0, A,...$ coefficients are, in principle, to be determined
phenomenologically. There is, however, an straightforward physical
interpretation for $B_0$. In the chiral limit, it is, up to a
normalization factor, the chiral condensate defined in eq.(\ref{defB_0}) 
(since it
yields a Lagrangian term which is proportional to the mass). That is:
 \be
\lim_{m_q\rightarrow0}\qqo=- 2 F^2 B_0 
\ee  

\subsubsection{Standard Chiral Perturbation Theory} 

As discussed above the standard scenario is to assume that $B_0$ is
large, $\Od(1\,\mbox{GeV})$, and thus it dominates the expansion in
eq.(\ref{masaycondensado}). Since $M_{meson}=\Opd$ we have that
$m_{quark}\simeq\Opd$, which is the standard chiral counting rule. If
we follow that counting, there is just one term that we can add to our
previous $\Opd$ chiral Lagrangian, which now becomes: \be {\cal
L}^{(2)}=\frac{F^2}{4}\left\{\tr(D_\mu UD^\mu U^\dagger)+
2B_0\tr({\cal M}(U^\dagger+ U))\right\} \ee Therefore, the meson
masses are given by: 
\bea 
M_\pi^2&=&2\hat{m}B_0\nn\\ M_K^2&=&(\hat{m}+m_s)B_0\nn\\ M_\eta^2
&=&\frac{2}{3}(\hat{m}+2m_s)B_0 \label{st_masses} 
\eea
 Just by cancelling the $B_0$ terms we get the Gell-Mann-Okubo mass relation, 
\be 
4M_K^2-M_\pi^2-3M_\eta^2=0 
\ee 
The fact that within this formalism this formula is obtained at
leading order, is an strong argument of plausibility for the initial
large $B_0$ assumption. We can also write the so called
Gell-Mann-Oakes-Renner formula in this notation as: 
\be
\frac{M_\pi^2}{2\hat{m}}-\frac{M_K^2}{\hat{m}+m_s}=
\frac{3M_\eta^2}{2(\hat{m}+2m_s)}=-\frac{\qqo}{2F^2} 
\ee 
In addition,
eq.(\ref{st_masses}), implies the following ratio of quark masses:

\be 
r\equiv\frac{m_s}{\hat{m}}=2\frac{M_K^2}{M_\pi^2}-1\simeq26  
\ee 
Indeed, the light quark masses are estimated using eq.(\ref{st_masses}),
which are obtained under the assumption that there is a large chiral
condensate. If the size of $\qqo$ turns out {\em not} to be that of
the standard scenario, then the current estimates of the
masses of the three lightest quarks \cite{udsmass} would have to be changed.
Concerning the
low energy theorems, within this standard formalism, they are modified
to:

\bea 
A(s,t,u)&=&\frac{s-M_\pi^2}{F^2}\nn\\ T^{3/2}(s,t,u)
&=&\frac{s-(M_\pi^2+M_K^2)}{2F^2} 
\eea 
Note, first, that now the meson elastic interaction does {\em not}
vanish at $s=0$. Indeed, {\em all the interaction at threshold is due
to the explicit chiral symmetry breaking}. Second, once we have
assumed the large condensate scenario the $\Opd$ predictions for the
amplitudes are {\em fixed}, since we know $F$, $M_\pi$ and $M_K$ from
experiment.  

We can now again build the most general $\Opc$ terms,
following the standard counting, and add them to the $\Opc$ Lagrangian
obtained in the previous section. That is, \cite{GaLe}
\bea 
{\cal L}^{(4)}&=&{\cal L}^{(4)}_{m_q=0}+ 2B_0L_4\; \tr(D^\mu U^\dagger
D_\mu U)\tr({\cal M}(U+U^\dagger))\nn\\ &&+ 2B_0L_5\,\tr(D^\mu
U^\dagger D_\mu U {\cal M}(U+U^\dagger)) + 4 B_0^2 L_6\,\tr({\cal
M}(U+U^\dagger))^2\nn\\ &&+4 B_0^2 L_7\,\tr({\cal M}(U-U^\dagger))^2
+4 B_0^2 L_8\,\tr({\cal M}U^\dagger{\cal M}U^\dagger+{\cal M}U{\cal
M}U)\label{stL4}
\eea  
This $\Opc$ Lagrangian now yields corrections to the pseudo-GB masses 
 which therefore modify the GMO formula to  \cite{GaLe}: 
\be 
4M_K^2-M_\pi^2-3M_\eta^2= \frac{6}{F^2}
(M_\eta^2-M_\pi^2)^2(L_5^r-6L_8^r-12L_7^r) 
-2(4M_K^2\mu_K-M_\pi^2\mu_\pi-3M_\eta^2\mu_\eta) 
\ee
where $\mu_a=(M_a^2/32\pi F^2)\log(M_a^2/\mu^2)$.  Unfortunately, the
correction depends on $L_8$ whose value cannot be determined with any
other independent experimental data. In that sense the Gell-Mann-Okubo
relation prediction does not ensure the dominance of the condensate
term in eq.(\ref{masaycondensado}).

\subsubsection{Generalized Chiral Perturbation Theory}

As we have just seen, there is still room for an alternative
scenario. Namely, to have a small condensate which therefore implies a
different counting scheme. In the extreme case, when $B_0=0$, we would
have $m_q\sim\Od(p)$, and this can be generalized to non-vanishing
small condensates with $B_0\sim\Od(p)$.  Hence, {\em although
we can still build the very same terms, now they will count differently} 
and their
relative importance at low energies will be modified. For instance, at
$\Opd$ we find \cite{GChPT1}:

\bea
\tilde{\cal L}^{(2)}&=&\frac{F^2}{4}\left\{\tr(D_\mu UD^\mu
U^\dagger)+ 2B_0\tr({\cal M}(U^\dagger+ U))\right.\nn\\
&+&A_0\,\tr({\cal M}U^\dagger{\cal M}U^\dagger+{\cal M}U{\cal M}U)
+Z_0^S\,\tr({\cal M}(U+U^\dagger))^2\nn\\ &+& \left. Z_0^P\,\tr({\cal
M}(U-U^\dagger))^2 + 2H_0\tr({\cal M}^2)\right\}\label{GChPTlag} 
\eea
and it can be noticed that the $A_0$, $Z_0^S$, and $Z_0^P$ terms have
the same structure as those of $L_8$, $L_6$ and $L_7$,
respectively. As a consequence of the new counting we now have, to
$\Opd$, the following masses:

\bea 
M_\pi^2&=&2\hat{m}B_0+4\hat{m}^2 A_0+4\hat{m}(2\hat{m}+m_s)Z_0^S\nn\\
M_K^2&=&(\hat{m}+m_s)B_0+(\hat{m}+m_s)^2A_0+2(\hat{m}+m_s)
(2\hat{m}+m_s)Z^S_0\nn\\
M_\eta^2&=&\frac{2}{3}(\hat{m}+2m_s)B_0+\frac{4}{3}(\hat{m}^2+2m_s^2)A_0\nn\\
&+&\frac{4}{3}(\hat{m}+2m_s)(2\hat{m}+m_s)Z^S_0+
\frac{8}{3}(m_s-\hat{m})^2Z^P_0 \label{g_masses} 
\eea
The appearance of so many new constants at leading order implies much
less predictive power. Indeed, it is not possible to obtain the
Gell-Mann-Okubo formula at this order since now 
\be
4M_K^2-M_\pi^2-3M_\eta^2=-4(\hat{m}-m_s)^2(A_0+2Z_0^P) 
\ee 
and
$(A_0+2Z_0^P)$ is not expected to vanish \cite{moreG}. But of course,
that does not mean that there is any incompatibility with experimental
data. 

Note that, using eq.(\ref{g_masses}), the quark mass ratio now ranges in
the interval
 \be 
r_1\equiv 2\frac{M_K}{M_\pi}-1\leq r \leq
2\frac{M_K^2}{M_\pi^2}-1 \equiv r_2 
\ee
That is: $6.3\leq r\leq26$, where 
the upper bound corresponds to
standard ChPT with $A_0=0$. Once more, the low energy
theorems are modified to \cite{GpiK}

\bea
 A(s,t,u)&=&\frac{\alpha_{\pi\pi}}{3F^2}M_\pi^2+
 \frac{\beta_{\pi\pi}}{F^2}\left(s-\frac{4}{3}M_\pi^2\right) \\
 T^{3/2}(s,t,u)&=&\frac{\alpha_{\pi K}}{3F^2}M_\pi M_K+
 \frac{\beta_{\pi K}}{4F^2}
 \left(t-\frac{2}{3}M_\pi^2-\frac{2}{3}M_K^2\right)+ \frac{\gamma_{\pi
 K}}{4F^2}(s-u)+ \frac{(M_K-M_\pi)^2}{6F^2}\nn
\eea 
where 
\bea
 \alpha_{\pi\pi}&=&1+6\frac{r_2-r}{r^2-1}\left(1+2\frac{Z_0^S}{A_0}\right)
 \quad;\quad \beta_{\pi\pi}=1\\ \alpha_{\pi
 K}&=&1+6\frac{r_2-r}{(r_1+1)(r-1)}\left(1+2\frac{Z_0^S}{A_0}\right)
 \quad;\quad \beta_{\pi K}=\gamma_{\pi K}=1
\eea 
In contrast with standard ChPT, we now deal with more parameters and
therefore {\em it is not enough to know} $F$, $M_\pi$ and $M_K$ to
determine the GChPT $\Opd$ predictions for the amplitudes
\cite{moreG}. Up to the moment we have limited the analysis of
Generalized Chiral Perturbation Theory to the $\Od(p^4)$
level. Following closely the philosophy of standard ChPT, it is also
possible to write down higher order Lagrangians, whose parameters can
absorb the infinities that do appear when computing loop
diagrams.

However, the counting scheme has been modified, and we have
introduced quantities that count as $\Od(p)$, therefore, we will now
have terms in the Lagrangian whose order in $p$ is an {\em odd
number}. In general, the GChPT Lagrangian is built of terms like
\cite{GChPT1,Gtwoloop}

\be
\tilde{\cal L}^{(d)}=\sum_{k+l+n}B_0^n{\cal L}_{(k,l)}, 
\quad \mbox{with}\quad{\cal L}_{(k,l)}\sim \Od(p^k m_q^l)
\ee
Indeed we have already given $\tilde{\cal L}^{(2)}$ in
eq.(\ref{GChPTlag}), and it is easy to see that $\tilde {\cal
L}^{(4)}_{(4,0)}$ is the very same ${\cal L}^{(4)}$ Lagrangian of
standard ChPT, eq.(\ref{stL4}). For the precise expressions of the rest
of the terms we refer to \cite{Gtwoloop}, but let us notice that we
have a Lagrangian at $\Od(p^3)$, which is not present in the standard
formalism. Only the $\tilde {\cal L}^{(4)}_{(2,2)}$ and$\tilde {\cal
L}^{(4)}_{(0,4)}$ terms have more than $18$ parameters and consequently,
the predictive power of the Generalized approach is somewhat weaker
than within the standard scenario. Nevertheless, it has been possible
to perform calculations, for instance, for the $\pi\pi$ scattering
lengths up to two loops, where only six different combinations of
chiral parameters appear in the final expressions \cite{Gtwoloop}.

\section{Present Evidence}

 Up to now we have just sketched the very basic consequences for the
 effective chiral formalism, of having either a large or small chiral
 condensate. Let us now review what support from data or other sources
 both alternatives have. 

The generalized scenario was motivated by some
 deviations in the $\pi N$ Goldberger-Treiman relation. For some
 values of the $\pi N$ coupling constant $g_{\pi N}$, it even
 suggested a ratio $r\simeq10$ \cite{Gold-Tre}, twice less than
 expected in the standard scenario, although the discrepancy was
 smaller for other values. More accurate data for 
 the $\pi N$ coupling constant before extracting any final
 conclusion. On the theoretical side, there are also some calculations
 which seem to prefer a lower value of the chiral condensate, like a
 variationally improved perturbation theory \cite{variation}, and some
 relativistic many-body approaches \cite{Cotanch}. Their low
 results, however, could be due to the approximations of their
 respective formalisms. One may think that the  determinations
 of light quark mass ratios could shed some light on this
 issue. However, due to the ambiguity pointed out in \cite{Manohar}
these ratios can only be obtained at lowest order using the standard ChPT
counting with the large condensate assumption.
In order to obtain the masses themselves
 some additional dynamical information has 
to be combined with chiral symmetry. At least,
it is possible to obtain a set of light quark masses consistent
with ChPT \cite{Leut}). 
The recent result $m_s(1
 \mbox{GeV})=235^{+35}_{-42}\,\mbox{MeV}$ reported by ALEPH
 \cite{ALEPH}, is consistent with both approaches, although the
 relatively high central value would be preferred by GChPT.

Concerning
 the standard scenario, we have seen that it was inspired by the
 Gell-Mann-Okubo and Gell-Mann-Oakes-Renner formula, which emerges
 very naturally at first order. In addition, the large condensate
 assumption receives a strong support from lattice calculations. Let
 us remember that within standard ChPT the pion mass {\em squared} is
 proportional to the quark masses.  That is indeed found in several
 lattice calculations \cite{lattice} which show a linear dependence of
 $M^2_\pi$ on $\mh$. This results have to be interpreted carefully
 since they involve an extrapolation, a quenched approximation and
 possible problems with the breaking of chiral symmetry when implementing
fermions on a lattice.\footnote{When we had finished this review we 
became aware of two new works related to this subject that give
more support to the standard scenario. Another work on the
lattice \cite{lattice2} seems to confirm previous results with 
better statistics. On ref.\cite{Shifman}
it is argued that, in the chiral limit, the low condensate scenario 
is ruled out by
QCD inequalities. However, they conclude that 
``one cannot rigurously rule out'' the alternative scenario
using QCD inequalities `` at finite quark mass'' although
``the condensate cannot be too small in order the QCD inequalities
to be satisfied at all distances''. } In addition, using
the hypothesis of global QCD-hadron duality, a QCD
sum rule  technique yields values for the condensate which 
support the standard framework \cite{Ximo}.

Thus there are no
 significant deviations in the data in conflict with either the
 standard or the generalized formalism, and the theoretical evidence
 always contains some approximation or extrapolation that obscures a
 conclusive statement. Fortunately it is very possible that the
 definitive answer will be obtained from experiment, by a precise
 measurement of the $\pi\pi$ scattering lengths. As we have seen in
 previous sections their non-vanishing values are exclusively due to
 the explicit symmetry breaking pattern. Whereas the large condensate
 scenario leads to a very sharp prediction of their values, the
 generalized framework allows for a wider range of values. Once again,
 the present experimental values of the scattering lengths are
 somewhat higher than expected within the standard formalism, but not
 in a strong disagreement.

\begin{table}
\begin{center}
\begin{tabular}{|c|c|c|c|c|c|}
\hline \hline
 & $ LET $ & $  ChPT (1l)$ &
 $ ChiPT (2l) $ & $ GChPT  (2l)$ &
$Exp$
 \\ 
\hline
 $ a_{00} $ & $0.16$ & $0.20 $ & $0.217$   & $0.263   $ 
& $0.26\pm 0.05   $\\
 $b_{00}$& $0.18$  & $0.26$ & $0.275$    & $0.25    $ 
& $0.25\pm 0.03    $\\
 $a_{20}$& $-0.045$& $-0.045$ &$-0.0413$ & $-0.027  $ 
& $-0.028\pm 0.012    $\\
 $b_{20}$& $-0.089$& $-0.085$ & $-0.072$ & $-0.079  $ 
& $-0.082\pm 0.008   $\\
 $a_{11}$& $0.030 $& $0.037$ & $0.040$   & $0.037   $ 
& $0.038\pm 0.002    $\\
 $b_{11}$& $- $    & $0.043$ & $0.0079$  & $0.054   $ 
& $ -   $\\
\hline \hline
\end{tabular}
\end{center}

\leftskip 1cm
\rightskip 1cm

{\footnotesize {\bf Table 1}
Scattering lengths $a_{IJ}$ and slope parameters $b_{IJ}$
for different $\pi\pi$ spin and isospin channels.
We show the predictions coming from the Low Energy Theorems (LET),
one-loop ChPT, two-loop ChPT and the two-loop GChPT {\em fit} 
to the scattering lengths, whose experimental values are given 
on the last column.}

\leftskip 0cm
\rightskip 0cm

\end{table}

In Table 1 we have listed the different values of the $\pi\pi$
scattering lengths, $a_{IJ}$ and slope
parameters $b_{IJ}$. In the last column we show the present experimental
values \cite{exp_lengths}.  Note that the errors are rather
large. The second column corresponds to the
$\Od(p^2)$ predictions in the standard formalism (Low Energy Theorems).
 Columns third and fourth correspond to the values
predicted by ChPT at $\Opc$ and $\Od(p^6)$ \cite{twoloop}.
Surprisingly, the $\Opc$ corrections can be as large as $20\%$ in some
cases. Since the values of $a_{00}$ and $a_{20}$ are above these
predictions, there was a hope that the two loop $\Opd$ corrections
would also turn out to be large and get a better result. Although they
contribute in the appropriate direction, they are not enough to get
the measured central value.  

The fifth column has been obtained using
GChPT at $\Od(p^6)$ \cite{Gtwoloop}.  As we have already mentioned, 
in this case, the predictions are not unique, since
they depend on the values of $\alpha$ and $\beta$. Thus, now it is possible
to include the scattering lengths as input in the fits.
As a matter of fact
the values given on the table are obtained from a combined fit 
to low energy experimental $K_{l4}$ decays {\em and} the 
phase shifts themselves, and they correspond to 
$\alpha =2.16$ and $\beta=1.074$ \cite{Gtwoloop}.
   
\section{The Chiral Phase Transition}

 Before discussing future experiments, let us briefly comment how the
 critical temperature of the Chiral Phase Transition could be changed
 if the standard scenario has to be modified.

In Fig.1 we show the
 evolution of the chiral condensate at finite $T$
 \cite{yo_mismo}. These results have been obtained studying the
 thermodynamics of a pion gas using the virial expansion.  In the
 second virial coefficient the $\pi\pi$ elastic scattering phase shifts
 are introduced to take into account the interactions of the
 pions. Thus, there are two effects that can modify the temperature
 dependence: first, obviously, the value of $\qqo$ itself, which sets
 the starting point at $T=0$. Second, the $\pi\pi$ interaction, which
 is different depending on whether we use the standard or generalized
 scenario. In Fig.1 we therefore see the interplay of both effects.

\begin{figure}[htb] 

\vspace*{1.0cm} 

\centerline{\epsfxsize=8cm \epsfbox{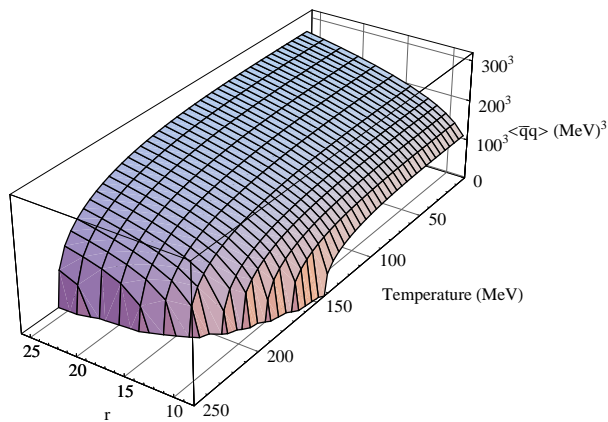}	
}  

\vspace{.5cm}

\caption[]{ Quark condensate versus temperature 
and $r$. } 
\label{fig1} 
\end{figure}
Note that at $r\simeq 24$ the generalized scenario reproduces
approximately the standard case. The condensate is of the order of
$(-250 \,\mbox{MeV})^3$and the critical temperature seems to be around
$200\, \mbox{MeV}$. When the higher order corrections are included in
GChPT, the minimum value of $r$ which is allowed is shifted from
$r_1\simeq 6.3$ to $r_1^*\simeq 8$ \cite{Gtwoloop}. That is why Fig.1
starts at $r=8$, which can therefore be interpreted as the extreme
non-standard case. All these results have to be taken cautiously,
since although the virial expansion could be a very good approximation
at those energies, we are neglecting the presence of more massive
states in the gas. An equivalent study within standard ChPT, showed
that massive states will spoil the approximation at about 150
MeV. However, their effect will always {\em decrease} the critical
temperature. In addition, the plot in Fig.1 corresponds to the central
values of all the parameters in GChPT, and some errors bands should
also be taken into account (For a more complete analysis we refer to
\cite{yo_mismo}). Nevertheless, the figure can illustrate how the non
standard scenario would predict that the condensate could melt at
considerably lower temperatures than in the standard picture. In the
extreme case, even below 150 MeV. That could have important
consequences in the study of quark gluon plasma and heavy ion physics.

\section{The DIRAC experiment}

In order to bring some light on the above discussed issues it was
proposed and now is under construction at CERN the DIRAC
\cite{DIRAC} experiment. 
By using the $24 \, \mbox{GeV}$ proton beam of the CERN Proton
Synchrotron this experiment will produce, among other things, pionia,
i.e. $\pi^+\pi^-$ atoms bounded by electromagnetic forces (first observed in
\cite{Afan93}). Those relativistic atoms $(\gamma \simeq 10)$ will either
decay in $\pi^0\pi^0$ pairs, get excited or be ionized. In the last
case characteristic charged pions called atomic pairs will emerge. The
two pions in these pairs have low relative momentum in their center of
mass system, small opening angle and nearly identical energies in the
lab system. 

The DIRAC experiment will consist on a double arm magnetic
spectrometer which will measure the number of atomic
pairs above the background of pion pairs produced in free states. For
a given target, the ratio of the number of atomic pairs to the total
number of pions depends on the pionium lifetime in
a unique way. Indeed, a simple
calculation shows that the lifetime $\tau_{n,0}$ of a pionium atom state 
with principal quantum number $n$ and orbital angular momentum $l=0$,
 can be written as
\be
\frac{1}{\tau_{n,0}} =  \frac{8\pi}{9}   
\sqrt{\frac{2\Delta m}{\mu}}\frac{(a_{00}-a_{20})^2}{M_{\pi}^2}
\vert \psi_{n,0}(0)      \vert^2    
\ee 
where  $\Delta m=
M-2M_{\pi^0}$, $M$ is the pionium mass, $\mu$ is its reduced mass and
$\psi_{n,0}$ is its wave function in the coordinate representation.
For Coulombian wave functions $\tau_{n,0}=\tau_{1,0}n^3$. For odd $l$
the decay of the pionium in a pair of neutral pions is forbidden and for
even $l>0$ this annihilation is suppressed by the square of the fine
structure constant. Therefore pionium annihilation takes place
mainly from $l=0$ states. That is why the total pionium lifetime happens to
be proportional to $(a_{00}-a_{20})^2$. 

The DIRAC experiment is expected to achieve a $10\%$
precision in the lifetime measurement, whose actual value 
is of the order of $10^{-15}s$. Thus the above 
combination of scattering lengths will be
determined to $5\%$. It is hoped that such resolution
would be enough to distinguish between the standard and the generalized
scenarios.

Further development of the DIRAC experiment could also lead to other
interesting measurements. For instance, strong interactions yield a
splitting of the energy levels $ns$ and $np$. In particular it can be
found \cite{EFIM86} that
\be 
\Delta E_{2s-2p}=-\frac{2\pi}{3M_{\pi}}\vert \psi_{1,0}(0) 
     \vert^2 (2a_{00}+a_{20})\simeq -0.3 \, \mbox{eV}
\ee

Therefore, a simultaneous measurement of the pionium lifetime and the
strong Lamb shift would be able to separate $a_{00}$ from $a_{20}$ in a
model independent way.

In addition, with some modifications, DIRAC may
produce $\pi K$ atoms and measure their  lifetime, which
is given by 
\be
\frac{1}{\tau_{n,0}} =  \frac{8\pi}{9}   
\sqrt{\frac{2\Delta m}{\mu}}(b_{1/2}-b_{3/2})^2\vert \psi_{n,0}(0)    
  \vert^2    
\ee 
where $\Delta m= M- M_{\pi_0}-M_{k_0}$, $M$ is the $\pi K$ 
atom mass and $\mu$ is the corresponding reduced mass.
Unfortunately,  in this case the
 relevant $\pi K$ scattering length combination ($b_{1/2}-b_{3/2}$)
is {\em not} sensible to the size of the condensate
\cite{GpiK} (note the different unit convention for the scattering lengths),
although it could be relevant to measure chiral parameters
like $L_1$, as suggested in  \cite{GaLe}.
Nevertheless, the present experimental data at threshold
are very bad \cite{exp-piK} and a better measurement would be
interesting to see how well ChPT works at those higher energies. 
As a matter of fact, it could be used
to test possible unitary extensions of the chiral approach like
the Inverse Amplitude Method \cite{IAM} which,
by using ChPT complemented with dispersion relations, makes it
possible to extend the energy range of applicability of plain ChPT. In
particular it describes quite well the available $\pi K$ scattering
data and it would be interesting to compare its predictions with the
more accurate data that could come from DIRAC.

What about KK-omnia? The production of such atoms has also been
studied for DIRAC \cite{mesatoms}.  Indeed that reference is a
complete study of the whole set of $\pi\pi$, $\pi K$ and $KK$
atoms. However, at such a high energies, it is necessary
to unitarize the ChPT predictions, for
instance by using the IAM method in a coupled channel formalism 
\cite{JA}. Nevertheless, the analogous treatment within GChPT 
is still to be done.

\section{Conclusions}

In this paper we have briefly reviewed the current status of the quark
condensate from the point of view of Chiral Effective Lagrangians.  
We have paid attention both to the theoretical and the
phenomenological role of this magnitude for large and small condensate
scenarios including the relevance for future heavy ion
experiments. We have seen that a value smaller than the standard one
for this magnitude is not ruled out by the present experimental
evidence. In particular we have reviewed how the standard Chiral
Perturbation Theory can be generalized in order to include the
possibility of a small condensate giving rise to the so called
Generalized Chiral perturbation Theory. 
That scenario could modify the standard estimates of
the critical 
temperature at which the chiral phase transition takes place.
Finally we have commented on the
DIRAC experiment which will provide important data in order to clarify
the issue of the precise value of the quark condensate and other
related with the unitarization of the ChPT amplitudes.

\section*{Acknowledgments} 

A.D. would like to thank Julia Nyiri for
giving him the opportunity to know better, in his very last days,
the great man that Gribov was and, at the same time, to meet a great
woman. J.R.P. thanks B. Adeva for suggesting him the
preparation of the lectures that lead to part of this review and
for explaining him in detail the DIRAC experiment, 
as well as J. Prades for his
explanations of sum rules and quark masses. He would also like to
thank the Theory Group at SLAC for their kind hospitality and the
Spanish Ministerio de Educaci\'on y Cultura for a Fellowship. This
work has been partially supported by the Spanish CICYT under
contract AEN93-0776 and by the U.S. Department of Energy under
contract DE-AC03-76SF00515.

\vfill\eject 

\begin{thebibliography}{99} 

\footnotesize
  
\bibitem{GOR}  M. Gell-Mann, Caltech Report CTSL-20 (1961).\\  
S. Okubo, {\em Prog. Theor. Phys.} {\bf 27} (1962) 949.\\ 
M. Gell-Mann, R.J. Oakes, and B. Renner, PR{175} (1968) 2195. 

\bibitem{GeLe}  P. Gerber and H. Leutwyler, {\it Nucl. Phys.}  {\bf B321},
  (1989) 387
 
\bibitem{QSR} M. Dey, V. L. Eletsky and B. L. Ioffe, \PL{252} (1990) 620\\
V. L. Eletsky and B. L. Ioffe, \PR{D47} (1993) 3083; \PR{D51} (1995) 2371\\
T. Hatsuda, Y. Koike and S.H. Lee, {\em Nucl. Phys.} {\bf B394} (1993) 221\\ 
G. Chanfray, M. Ericson and J. Wambach, {\em Phys. Lett.} {\bf B388} (1996) 673

\bibitem{We} S. Weinberg, {\em Physica} {\bf A} (1979) 327. 

\bibitem{GaLe} J. Gasser and H. Leutwyler,  
Ann. of Phys. {\bf 158}, (1984) 142, \NP{B250} (1985) 465.

\bibitem{GChPT1} N. H. Fuchs, H. Sazdjian and J.Stern, 
\PL{B269} (1991) 183.\\  
J.Stern, H.Sazdjian and N.H. Fuchs, \PR{D47} (1993) 3814. 

\bibitem{Stern} J. Stern, hep-ph/9801282. 

\bibitem{libro} A. Dobado, A. G{\'o}mez-Nicola, 
A. L. Maroto and J. R. Pel\'aez. 
{\em Effective Lagrangians for the Standard Model}. 
Texts and Monographs in Physics. Springer-Verlag (1997). 
 
\bibitem{lattice} S. Aoki et al. (CP-PACS)   {\it Nucl. Phys (Proc. Suppl.)} 
{\bf B60A} (1998) 14.\\  V. Gim\'enez et al., {\em hep-lat/9801028}  

\bibitem{Leut} H. Leutwyler,  \NP{A623} (1997) 169c. (hep-ph/9709406)

\bibitem{ALEPH} S. Chen (for the ALEPH Collab.), 
{\em Nucl. Phys. (Proc. Suppl.)} {\bf 64} (1998) 265.

\bibitem{Gold-Tre}  N. H. Fuchs, H. Sazdjian and J.Stern, \PL{B238} (1990) 380.

\bibitem{variation}  G. Arvanitis et al., \PL{B390} (1997) 385.\\
J.L. Kneur, \PR{D57} (1998) 2785; {\it Nucl. Phys (Proc. Suppl.)}
 {\bf B64} (1998) 296.

\bibitem{Cotanch} A. Szczepaniack et al., \PRL{76} (1996) 2011. 
 
\bibitem{Manohar} D.B. Kaplan and A. V. Manohar, \PRL{56} (1986) 2004.

\bibitem{Ximo} J. Bijnens, J. Prades and E. de Rafael, 
\PL{B348} (1995) 226.\\ J. Prades,  {\it Nucl. Phys (Proc. Suppl.)} 
{\bf B64} (1998) 253.

\bibitem{moreG} M. Knecht and J. Stern, {\em The Second DA$\Phi$NE
 Physics  Handbook}, 
eds: L. Maiani, G. Pancheri and N. Paver., INFN, Frascati (1995)  
(hep-ph/9411253).\\  
J. Stern, {\em Nucl. Phys. (Proc. Suppl.)}{\bf 64} (1998) 232.  

\bibitem{GpiK} M. Knecht, H. Sazdjian, J.Stern and N. H. Fuchs, 
\PL{B313} (1993) 229.

\bibitem{twoloop} J. Bijnens et al. \PL{B374} (1996) 210;  
\NP{B508} (1997) 263.  

\bibitem{Gtwoloop} M. Knecht, B. Mousallam and J. Stern, \NP{B457} (1995) 513.

\bibitem{yo_mismo} J. R. Pel\'aez, SLAC-PUB 7865. hep-ph/9806532. 
     
\bibitem{exp_lengths} J. L. Basdevant, C. D. Froggat and J. L. Petersen, 
\NP{B72} 413 (1974)\\ 
J. L. Basdevant, P. Chapelle,  C. L\'opez and M. Sigelle, 
\NP{B98}, (1975) 285.\\ 
C. D. Froggat and J. L. Petersen, \NP{B129} (1977) 89.\\ 
J. L. Petersen, {\em The $\pi\pi$ interaction},  CERN Yellow Report No.77-04 
(1977)

\bibitem{udsmass} H. Leutwyler, \PL{B378} (1996) 313 (hep-ph/9602366)  

\bibitem{DIRAC} B. Adeva et al.  
{\em Lifetime measurement of $\pi^+\pi^-$ atoms to  
test low-energy QCD predictions}, 
Proposal to the SPSLC, CERN/SPSLC 95-1,  SPSLC/P 284, Geneva 1995.  

\bibitem{pipitheory} R. Lednicky and V. L. Lyuboshitz,  
{\em Yad. Fiz. }{\bf 35} (1982) 1316.\\  
L.L. Nemenov, {\em Yad. Fiz. }{\bf 41} (1985) 980.  

\bibitem{Afan93} L.G. Afanasyev et al., \PL{B255} (1991) 146.

\bibitem{EFIM86} G.V. Efimov, M.A. Ivanov and Lyubovitskij, 
{ Yad. Fiz.  {\bf 44}} (1986) 460 
   
\bibitem{IAM} A. Dobado and J.R. Pel\'aez, \PR{D56} (1997) 3073;
  \PR{D56} (1997) 3057. 
  
\bibitem{exp-piK} V. Bernard, N. Kaiser and U. G. Mei{\ss}ner, 
 \PR{D43} (1991) 2757  

\bibitem{Penn} M.R. Pennington, \NP{A623} (1997) 189c.  
 
\bibitem{mesatoms} S. Wycech and A. M. Green, \NP{A562} (1993) 446.  

\bibitem{JA} J.A. Oller, E. Oset and J.R. Pel\'aez, \PRL{80} (1998) 3452;
hep-ph/9804209.\\
F. Guerrero and J. A. Oller, hep-ph/9805334.

\bibitem{lattice2}L. Giusti, F. Rapuano, M.Talevi and A. Vladikas. 
hep-lat/9807014

\bibitem{Shifman} I.I. Kogan, A. Kovner and M. Shifman. hep-ph/9807286
\end{thebibliography}
\end{document}